\documentclass[11pt, a4paper]{article}
\usepackage{jheppub}

\usepackage{amssymb}
\usepackage{citesort}
\usepackage{amsthm}
\usepackage{amsfonts}
\usepackage[active]{srcltx}

.tfm scaled 1000 
.tfm scaled 1000 
.tfm scaled 1000 

\newcommand{\To}{\longrightarrow}

\title{Liouville theory, ${\cal N}=2$ gauge theories
and accessory parameters}

\author[a]{Franco Ferrari,}
\author[a,b]{Marcin Pi\c{a}tek}

\affiliation[a]{Institute of Physics and CASA*, University of Szczecin,\\
ul. Wielkopolska 15, 70-451 Szczecin, Poland}
\affiliation[b]{Bogoliubov Laboratory of Theoretical Physics,
Joint Institute for Nuclear Research,\\
141980 Dubna, Russia}

\emailAdd{ferrari@fermi.fiz.univ.szczecin.pl}
\emailAdd{piatek@fermi.fiz.univ.szczecin.pl}

\abstract{
The correspondence between the semiclassical limit of the DOZZ quantum
Liouville theory and the Nekrasov--Shatashvili limit of the ${\cal N} = 2$
($\Omega$-deformed) ${\sf U(2)}$ super--Yang--Mills theories is used to
calculate the unknown accessory parameter of the Fuchsian uniformization
of the 4-punctured sphere. The computation is based on the saddle point
method. This allows to find an analytic expression for the $N_f = 4$,
${\sf U(2)}$ instanton twisted superpotential and, in turn, to sum up the 4-point
classical block.
It is well known that the critical value of the Liouville
action functional is the generating function of the accessory parameters.
This statement and the factorization property of
the 4-point action allow to express the unknown accessory
parameter as  the derivative of the 4-point classical block with respect
to the modular parameter of the 4-punctured sphere.
It has been found that this accessory
parameter is related to the sum of all rescaled column lengths of the so-called
'critical' Young diagram extremizing the instanton 'free energy'.
It is shown that the sum over the 'critical' column lengths can be rewritten
in terms of a contour integral in which the integrand is built out of certain special
functions closely related to the ordinary Gamma function.}

\keywords{Supersymmetric gauge theory, Field Theories in Lower Dimensions, Solitons
Monopoles and Instantons}

\arxivnumber{1202.2149}

\begin{document}
\maketitle

\section{Introduction}
The studies of the interrelationships between conformal field theory
in two dimensions, supersymmetric ${\cal N}=2$ quiver gauge theories and
integrable systems are recently attracting a great attention
in the scientific community
\cite{NekraRosSha,BelavinBelavin,FatLitv,AlbaFatLitTarnp,Tai,Tai2,Pogho1,MarTaki,Teschner,NekraWitten,BonelliTanzini,Piatek}.
This is mainly due to the discovery of the so-called AGT \cite{AGT}
and Bethe/gauge
\cite{NekraSha,NekraSha2,NekraSha3} correspondences.

The AGT conjecture states that the Liouville field theory (LFT) correlators on the
Riemann surface $C_{g,n}$ with genus $g$ and $n$ punctures can be identified with the
partition functions of a class $T_{g,n}$ of four-dimensional ${\cal N}=2$ supersymmetric ${\sf SU}(2)$ quiver gauge
theories. A significant part of the AGT conjecture is an exact correspondence between the Virasoro
blocks on $C_{g,n}$ and the instanton sectors of the Nekrasov partition functions of the gauge theories
$T_{g,n}$. Soon after its discovery, the AGT hypothesis has been
extended to the $\sf SU(N)$-gauge
theories/conformal Toda correspondence \cite{Wyllard,MMU(3),MMMM}.

The AGT correspondence works at the level of the quantum Liouville
field theory. It arises at this point
the question of what happens if we proceed to the classical limit
of the Liouville theory. It turns out that the semiclassical limit of
the LFT correlation functions \cite{Zamolodchikov:1995aa}, i.~e. the limit in
which the central charge and the external and intermediate conformal weights tend to infinity
while their ratios are fixed, corresponds to the
Nekrasov--Shatashvili limit of the Nekrasov partition functions \cite{NekraSha}.
In particular, a consequence of that correspondence is that the {\it classical
conformal block} can be identified with the instanton sector of the
{\it effective twisted superpotential}
\cite{Piatek}.\footnote{
One can see that a slightly more general statement holds,
i.e. the so-called {\it classical Liouville action} \cite{Zamolodchikov:1995aa}
can be connected with the full twisted superpotential.} The latter
quantity determines the low energy effective
dynamics of the two-dimensional gauge theories restricted to the
$\Omega$-background. The twisted superpotential
plays also a pivotal role in the already mentioned Bethe/gauge
correspondence
that maps supersymmetric vacua of the ${\cal N}=2$ theories to Bethe
states of quantum integrable systems. A result of that duality is that
the twisted superpotential is identified with the Yang's functional
\cite{YY} which
describes the spectrum of the corresponding quantum integrable system.
Joining together the AGT duality and the Bethe/gauge
correspondence it is thus possible to link the classical blocks
(or more in general the classical Liouville actions) to the
Yang's functionals.

The motivations to study the classical block were until now mainly
confined to applications in pure mathematics, in particular to the
celebrated uniformization problem, which roughly speaking is related
to the construction of conformal mappings between Riemann surfaces (RS)
admitting a simply connected universal covering and the three existing
simply connected RS, the sphere, the complex plane and the upper half plane.
The uniformization problem is well illustrated by the example
of the uniformization of the Riemann sphere with four punctures \cite{HJuniformizacja}.
Its uniformization may be associated to a Fuchsian equation whose
form is known up to some constants that are called {\it accessory
parameters}. Their computation is an open longstanding problem, which can however be solved
if we succeed to derive an analytical expression of the classical block obtained
by performing the classical limit of the four-point correlation
function of the DOZZ quantum Liouville field theory \cite{Zamolodchikov:1995aa,DO}.
The importance of the classical blocks is not only limited to the
uniformization theorem, but gives also information about the
solution of the Liouville equation on surfaces with punctures.
For instance, if the accessory parameters
for $C_{0,4}$ are available, it is then possible to construct the solution of
the Liouville equation and the hyperbolic metric on  $C_{0,4}$.
Due to the recent discoveries mentioned above the classical blocks
have become  relevant also in mathematical and
theoretical physics, since they are related to quantum integrable
systems\footnote{
The classical conformal block and the accessory parameters have
fascinating interpretations in that
context. It has been recently found in \cite{NekraRosSha} that the
classical conformal block
corresponds to the generating function of the so-called variety of opers which
has been introduced to define the Yang's functional. In \cite{Teschner}
the accessory parameters have been identified with the Hitchin Hamiltonians.}
and to the instantonic sector of certain ${\cal N}=2$ supersymmetric
gauge field theories.\footnote{Surprising relationships between the LFT
and the ${\cal N}=2$ SYM theories were also observed before the
discovery of the AGT correspondence, see for instance \cite{MatoneAnd}.}

The link between classical blocks and Yang's functionals has been
exploited in \cite{Piatek}
to conjecture a novel representation of the 4-point classical block
in terms of the elliptic Calogero-Moser (eCM) Yang's functional found in
\cite{NekraSha}. As an application of that result
the relation between the accessory parameter
of the Fuchsian uniformization of the 4-punctured Riemann sphere
and the eCM Yang's functional has been proposed \cite{Piatek}.
However, the results described above have an important limitation.
They are not general, i.e. they hold only for certain classes
of classical block parameters or, in other words, for restricted
families of the 4-punctured spheres.

The purpose of the present paper is to find an analytical expression
of the {\it generic} classical 4-point block and apply it to compute the
unknown accessory prameter appearing in the Fuchsian differential
equation with four elliptic/parabolic singularities.  In order
to accomplish this task we will employ the
correspondence mentioned above between the classical limit of
the Liouville theory and
the Nekrasov--Shatashvili limit of the ${\cal N} = 2$
($\Omega$-deformed) ${\sf U(2)}$ super-Yang-Mills theories.
The relevant technical problem of
this strategy consists in the
summation of the series defining the twisted superpotential
(and/or the classical block). This problem will be tackled hereusing
the saddle point method \cite{Pogho1,N,NekraOkun,Fucito}.

The structure of the paper is as follows. In section 2 we formulate
the problem of the accessory parameters of the Fuchsian uniformization of the
$n$-punctured sphere and describe its connection with the classical Liouville theory.
Afterwards, we briefly review the so-called {\it geometric approach} to quantum Liouville
theory originally proposed by Polyakov
(as reported in refs. \cite{TZ,TZ2,Takhtajan:yk,Tak,ZoTa2})
and further developed by Takhtajan \cite{Takhtajan:yk,Tak,Takhtajan:1993vt,Takhtajan:zi}
(see also \cite{TakTeo}).
Some of the predictions derived from the path integral
representation of the geometric approach can be  proved rigorously and lead to
deep geometrical results. One of these results has been the
suggestion
that the classical Liouville action is the
generating function for the accessory
parameters of the Fuchsian uniformization of the punctured
sphere. This statement
yields an essentially new insight into the problem of accessory parameters.
However, its usefulness is restricted by our ability to calculate the classical
Liouville action for more than three singularities.
We focus on the case of the sphere with four punctures
and show that there is only one unknown accessory parameter whose
computation is equivalent to the calculation of the 4-point classical block.

In section 3 we compute the Nekrasov--Shatashvili limit of the Nekrasov
instanton partition function of the ${\sf U(2)}$, $N_{f}=4$, ${\cal N}=2$ SYM
theory closely following \cite{Pogho1} (see also \cite{Fucito}).
On the basis of arguments worked out by Nekrasov and Okounkov \cite{NekraOkun}
it is found that the effective
twisted superpotential is equal to the critical value of the instanton
'free energy'. The critical (or classical) instanton configuration is
determined by a saddle point equation that can be solved recursively
order by order in the instanton parameter $q$. In the language of Young
diagrams the solution of the saddle point equation describes the shape of
the most relevant 'critical' Young diagram contributing to the instanton
partition function. We check that the instanton 'free energy'
evaluated at the critical configuration (i.e. the twisted superpotential)
gives the correct $q$-expansion of the classical 4-point block provided that
certain relations between the parameters are holding. Taking these
relations  into
account, we are able to express
the 4-point classical block in terms of the twisted superpotential
and apply this representation to calculate the unknown accessory
parameter. We find in this way that the accessory parameter is related
to the
sum of all column lengths of the 'critical' Young diagram.
As it has been shown in \cite{Pogho1}, this sum can be rewritten as a
contour
integral where the integrand contains as an essential ingredient
certain special
functions closely related to the ordinary Gamma function (cf. \cite{Fucito}).
In Section 4 we present our conclusions. The problems that are still
open and the possible extensions of the present work are discussed.
Finally, in the appendix we define the quantum and classical four-point
conformal blocks.

\section{Liouville theory and accessory parameters}
\subsection{Monodromy problem and uniformization}
Let us choose a set of complex coordinates
$z_1,\ldots,z_n$
on the $n$-punctured Riemann sphere
$C_{0,n}$
in such a way that $z_{n}=\infty$. The so-called problem of accessory
parameters
can be formulated as follows. Consider the ordinary linear
differential equation:
\begin{eqnarray}\label{Fuchs}
\partial^{2}_{z} \psi(z) + T^{\,\sf cl}(z)\,\psi(z) \;=\; 0,
\end{eqnarray}
where $T^{\,\sf cl}(z)$ is a meromorphic function on the Riemann sphere of the form:
\begin{eqnarray}
\label{tensor1}
T^{\,\sf cl}(z) &=&
\sum_{k=1}^{n-1}\left[\frac{\delta_k}{(z - z_k)^2}+
\frac{c_k}{z - z_k}\right]
\end{eqnarray}
and
\begin{equation}
\label{tensor2}
T^{\,\sf cl}(z) =
\frac{\delta_n}{z^2} + \frac{c_n}{z^3} + {\cal O}\left(z^{-4}\right)
\hskip 1.5cm  {\rm for}\;\; z \to \infty
\end{equation}
with
$$
\delta_i = \frac{1}{4}(1-\xi_{i}^{2}),
\;\;\;\;\;\;\;\;\;
\xi_{i}\in \mathbb{R}_{\geq 0},
\;\;\;\;\;\;\;\;\;
i=1,\ldots, n.
$$
The asymptotic behaviour (\ref{tensor2}) of $T^{\,\sf cl}(z)$ implies
that the coefficients $c_1,\ldots,c_n$,
known as the accessory parameters, obey the relations
\begin{equation}
\label{dodatki2}
\sum_{k=1}^{n-1}c_k =0,
\;\;\;\;\;\;\;\;\;\;
\;\;\;\;\;\;\;
\sum_{k=1}^{n-1}\left(\delta_k + c_k z_k\right)=\delta_n,
\;\;\;\;\;\;\;\;\;\;
\;\;\;\;\;\;\;
\sum_{k=1}^{n-1}\left(2\delta_k z_k + c_k z_{k}^{2}\right)=c_n .
\end{equation}
The problem is to tune these parameters in such a way that
the eq. (\ref{Fuchs}) admits a fundamental system of solutions with
monodromy in ${\sf
PSL}(2,\mathbb{R})$.\footnote{Equivalently, one could employ
here the group ${\sf PSU}(1,1)$
which is isomorphic to ${\sf PSL}(2,\mathbb{R})$.}
Note that for $n>3$ the equations (\ref{dodatki2}) do not provide
enough constraints
in order to calculate all the $c_k$'s. The computation of the
accessory parameters in that case is difficult and in general still unsolved
problem.

One can shed some light on the form of the $c_k$'s
by relating the problem of finding the accessory parameters
to the Liouville field theory on $C_{0,n}$. Indeed, let us consider the quotient:
\begin{equation}
\rho = \frac{\psi_1}{\psi_2}\label{rhodef}
\end{equation}
of the fundamental solutions $(\psi_1, \psi_2)$ of the eq. (\ref{Fuchs})
with Wronskian $\psi_1 \psi_{2}' - \psi_{1}'\psi_{2}=1$ and
${\sf SL}(2, \mathbb{R})$ monodromy with respect to all punctures.
It is a well known fact \cite{Hempel,TZ} that
$$
\rho\! : C_{0,n}\ni z\To \tau(z)\in\mathbb{H}
$$
is a multi-valued map from the $n$-punctured Riemann sphere
to the upper half plane $\mathbb{H}=\lbrace \tau\in\mathbb{C}\,:\,\mathfrak{Im}\,\tau
> 0\rbrace$ with branch points $z_1, \ldots, z_n$.
The connection with Liouville theory  comes out from the existence of
the Poincar\'{e} metric
$
d\textrm{s}^{2}_{\mathbb{H}}=d\tau\,d\bar\tau/\left(
\mathfrak{Im}\,\tau\right)^2$
on the upper half plane $\mathbb{H}$. The pull back
\begin{equation}\label{pullback}
\rho^{\ast}d\textrm{s}^{2}_{\mathbb{H}}=
\frac{1}{\left( \mathfrak{Im}\,\tau\right)^2}\left|
\frac{\partial\tau}{\partial z}\right|^2 \ dz d\bar
z = \textrm{e}^{\phi(z,\bar z)} dz d\bar z
\end{equation}
is a regular hyperbolic metric on $C_{0,n}$, conformal to
the standard flat matric $dz d\bar z$ on
$\mathbb{C}$.
The conformal factor $\phi(z, \bar z)$ of that metric
satisfies the Liouville equation:
\begin{equation}
\label{Liouville}
\partial_z\partial_{\bar z} \phi(z,\bar z) =
\frac{1}{2}\, \textrm{e}^{\phi(z,\bar z)}
\end{equation}
and has one of the following asymptotic behaviors near the punctures:
\begin{enumerate}
\item
case of {\it elliptic singularities}:
\begin{eqnarray}
\label{asymptot:elliptic}
&&\hspace{-50pt}\phi(z,\bar z) = \left\{
\begin{array}{lll}
-2\left(1-\xi_j \right)\log | z- z_j |  + O(1) & {\rm as } & z\to
z_j,
\hskip 5mm j = 1,\ldots,n-1,\\
-2\left(1+\xi_n \right)\log | z| + O(1) & {\rm as } & z\to \infty,
\end{array}
\right.
\\[3pt]
\label{Picard}
&&\hspace{10pt}
\xi_i \in \mathbb{R}_{>0}\;\;\;\textrm{for}\;\textrm{all}\;\;i=1,\ldots,n
\;\;\;\textrm{and}\;\;\sum\limits_{i=1}^{n}\xi_i < n-2;\nonumber
\end{eqnarray}
\item
case of {\it parabolic singularities} ($\xi_i \to 0$):
\begin{equation}
\label{asymptot:parabolic} \phi(z,\bar z) = \left\{
\begin{array}{lll}
-2\log |z- z_j |  -2\log \left|\log |z- z_j |\right| + O(1) & {\rm as } & z\to z_j, \\
-2\log |z| - 2\log \left|\log |z|\right| + O(1) & {\rm as } & z\to
\infty.
\end{array}
\right.
\end{equation}
\end{enumerate}
It is known that it exists a unique solution of eq.~(\ref{Liouville})
if one of the conditions
(\ref{asymptot:elliptic})
\cite{Picard1,Picard2,Troyanov}
or (\ref{asymptot:parabolic}) \cite{Heins} is satisfied.
One may show that the meromorphic function $T^{\,\sf cl}(z)$
introduced in eqs.~(\ref{tensor1}), (\ref{tensor2}) is the holomorphic
component of the energy-momentum tensor:
\begin{equation}\label{Schwarz}
T(z) \equiv - \frac{1}{4}\,(\partial_{z}\phi)^2 +
\frac{1}{2}\,\partial_{z}^{2}\phi
\end{equation}
evaluated at the solution $\phi(z,\bar z)$ of the Liouville equation
with one of the asymptotic
conditions\footnote{In the case of parabolic singularities the {\it
    classical conformal weights} in
(\ref{tensor1}), (\ref{tensor2}) are $\delta_i =
  \frac{1}{4}\;\Longleftrightarrow\;
\xi_i \to 0$.}
(\ref{asymptot:elliptic}) or (\ref{asymptot:parabolic}).
Once the classical solution is known, it is possible to calculate all
the accessory parameters.

The monodromy problem for the Fuchs equation (\ref{Fuchs}) formulated
above has been proposed
by Poincar\'{e} in order to construct the so-called {\it uniformization map}
in the case of the $n$-punctured sphere with parabolic singularities.
To derive the uniformization map for the $n$-punctured sphere $C_{0,n}$
it is necessary
to compute a meromorphic map $\lambda$ from the upper half-plane $\mathbb{H}$
to $C_{0,n}$ such that $\lambda$ is the covering map
$$
\lambda:\mathbb{H} \To \mathbb{H}/G \simeq C_{0,n}
$$
with $G$ being a discrete subgroup of ${\sf PSL}(2,\mathbb{R})$.
The computation of $\lambda$ is the main problem of the Fuchsian
uniformization scheme.
If $\lambda$ is the uniformization map of $C_{0,n}$, then the
multi-valued function
$\rho$ defined in eq.~(\ref{rhodef}) coincides with the inverse map
$\lambda^{-1}$, i.~e. $\rho=\lambda^{-1}$. The
branches of $\rho$ are related by the
elements of the group $G$. It is possible to show that the classical
energy-momentum tensor
$T^{\,\sf cl}(z)$ is equal to one half of the Schwarzian derivative of
the map $\rho$:
\begin{equation}\label{Schwarz2}
T^{\,\sf cl}(z) = \frac{1}{2}\,\lbrace\rho, z\rbrace.
\end{equation}
Thus, the inverse map $\rho$ can be computed if the appropriate solution of the
Liouville equation is available or, equivalently, if the accessory parameters
in the Fuchs equation (\ref{Fuchs}) are known and one may select
fundamental solutions of it with
a suitable monodromy.

In the case of elliptic singularities the multi-valued function $\rho$ is also
of interest. It is no longer the inverse to the covering map of  $C_{0,n}$, but
it can still be used to construct solutions of the Liouville equation with
the asymptotic behavior (\ref{asymptot:elliptic}) according to the
formula (\ref{pullback}).
Thus, the result (\ref{Schwarz2}) holds also when the singularities
are elliptic and, as before, the problem
of calculating the map $\rho$ is equivalent to that of finding
the solution of the Liouville equation or solving the monodromy
problem for the Fuchs equation (\ref{Fuchs}).

\subsection{Liouville action and accessory parameters}
For almost a century the problem of accessory parameters has remained
unsolved until the appearance of its solution proposed by Polyakov
(as reported in refs. \cite{TZ,Tak,TZ2}).
Polyakov observed that the (properly defined
and normalized) Liouville action functional evaluated at the
classical solution $\phi(z, \bar z)$ is the generating functional for
the accessory parameters:
\begin{equation}
\label{PC} c_j = -\frac{\partial S_{\sf L}^{\,\sf
cl}[\phi]}{\partial z_j}.
\end{equation}
This formula was derived within the so-called {\it geometric path
  integral approach} to the
quantum Liouville theory by analyzing the quasi-classical limit of the
conformal Ward identity \cite{Tak}.

In the geometric approach
the correlators of the LFT are expressed in terms of path integrals
over the conformal class of Riemannian metrics
with prescribed singularities at the punctures. In particular, in the case
of the quantum Liouville theory on the sphere the central objects
in the geometric approach are\footnote{In this
subsection we closely follow \cite{Takhtajan:yk}.}
\begin{enumerate}
\item
the 'partition functions' on $C_{0,n}$:
\begin{equation}
\label{partition}
\left\langle\; C_{0,n} \;\right\rangle =
\int\limits_{\cal M}{\cal D}\phi\;{\rm e}^{-Q^2 S_{\sf L}[\phi]},
\end{equation}
where $\cal M$ is the space of conformal factors appearing in the
metrics on $C_{0,n}$ with either the
asymptotic behavior of
eq.~(\ref{asymptot:elliptic}) or that of eq.~(\ref{asymptot:parabolic});
\item
the correlation functions of the energy-momentum tensor:
\begin{eqnarray}
\label{emom}
\left\langle
\widehat{T}(u_1)\ldots \widehat{T}(u_k) \widehat{\bar T}(\bar w_1)
\ldots \widehat{\bar T}(\bar w_l)\, C_{0,n}
\right\rangle  &=&\nonumber
\\
&&\hspace{-100pt}=\;
\int\limits_{\cal M}\!\!{\cal D}\phi\;{\rm e}^{-Q^2 S_{\sf L}[\phi]}\;
\widehat{T}(u_1)\ldots \widehat{T}(u_k) \widehat{\bar T}(\bar w_1)
\ldots \widehat{\bar T}(\bar w_l)
\end{eqnarray}
with
\begin{equation}
\label{emom:def}
\widehat{T}(u) = Q^2\left[
-\frac{1}{4}\left(\partial_u\phi(u,\bar u)\right)^2 + \frac{1}{2}\,\partial^2_u\phi(u,\bar u)
\right].
\end{equation}
\end{enumerate}

The singular nature of the Liouville field at the punctures
requires regularizing terms in the Louville action:
\begin{eqnarray}
\label{action}
S_{\sf L}[\phi]
& = &
\frac{1}{4\pi}
\lim_{\epsilon\to 0}
S_{\sf L}^\epsilon[\phi]
\end{eqnarray}
where the regularized action $S_{\sf L}^\epsilon[\phi]$ is given by
\begin{eqnarray}
S_{\sf L}^\epsilon[\phi]
& = &
\int\limits_{X_\epsilon}\!d^2z
\left[\left|\partial\phi\right|^2 + {\rm e}^{\phi}\right]
+  \sum\limits_{j=1}^{n-1}\left(1-\xi_j\right)
\hspace{-4mm}\int\limits_{|z-z_j|=\epsilon}\hspace{-4mm}|dz|\ \kappa_z \phi
+\left(1+\xi_n\right)
\hspace{-2mm}\int\limits_{|z|=\frac{1}{\epsilon}}\hspace{-2mm}|dz|\ \kappa_z
\phi\nonumber
\\
&&
- 2\pi\sum\limits_{j=1}^{n-1}\left(1-\xi_j\right)^2\log\epsilon
- 2\pi\left(1+\xi_n\right)^2\log\epsilon,\label{regLFTaction}
\end{eqnarray}
and \( X_\epsilon = {\mathbb C}\setminus\left\{\left(\bigcup_{j=1}^n |z-z_j|< \epsilon\right)
\cup \left(|z|>\frac{1}{\epsilon}\right)\right\}. \) The prescription
given in
eqs.~(\ref{action}) and (\ref{regLFTaction})
is valid for parabolic singularities (corresponding to
$\xi_j = 0$) as well.

One can check by perturbative calculations of the correlators (\ref{emom})
\cite{Takhtajan:yk} that
 the central charge reads
\begin{equation}
\label{centrsl:charge}
c = 1 + 6Q^2.
\end{equation}
The transformation properties of (\ref{partition}) with respect to global conformal
transformations show \cite{Takhtajan:yk} that the punctures behave as
primary fields with dimensions
\begin{equation}
\label{Delta}
\Delta_j =
\bar{\Delta}_j = \frac{Q^2}{4}\left(1-\xi_j^2\right).
\end{equation}
For fixed $\xi_j$, the dimensions scale like $Q^2$ and
the punctures correspond to {\it heavy} fields
of the operator approach \cite{Zamolodchikov:1995aa}.
In the classical limit $Q^2 \to \infty$ with all classical weights
\begin{equation}
\label{small:deltas}
\delta_i  \stackrel{\rm def}{=}   \frac{\Delta_i}{Q^2} = \frac{1-\xi_j^2}{4}
\end{equation}
kept fixed,  we expect the path integral to be
dominated by the classical action $S^{\,\sf cl}_{\sf L}(\delta_i\,;\,z_i)$,
\begin{equation}
\label{asymptotic:X}
\left\langle \; C_{0,n} \; \right\rangle
\sim
{\rm e}^{-Q^2S^{\,\sf cl}_{\sf L}(\delta_i\,;\,z_i)}.
\end{equation}
In the above equation $S^{\,\sf cl}_{\sf L}(\delta_i\,;\,z_i)$ denotes the
functional $S_{\sf L}[\,\cdot\,]$ of eq.~(\ref{action}) evaluated at the
classical solution
 $\phi$  of (\ref{Liouville}) with the asymptotics
(\ref{asymptot:elliptic}) or (\ref{asymptot:parabolic}).
A similar result holds for the correlation function $\left\langle
\widehat{T}(z)\;C_{0,n} \right\rangle$:
\begin{equation}
\label{classical:withT}
\left\langle \widehat{T}(z)\;C_{0,n} \right\rangle
\sim
\widehat{T}^{\,\sf cl}(z)\ {\rm e}^{-Q^2 S^{\,\sf cl}_{\sf L}(\delta_i\,;\,z_i)},
\end{equation}
where $\widehat{T}^{\,\sf cl}(z)$ is the classical energy-momentum tensor.

From (\ref{emom:def}) and (\ref{asymptot:elliptic}) or
(\ref{asymptot:parabolic}) it follows that
\begin{eqnarray}
\label{singular:behavior}
\nonumber
\widehat{T}^{\,\sf cl}(z) & \sim &  \frac{\Delta_j}{(z-z_j)^2}
 \hskip 1cm  {\rm for}\;\; z \to z_j, \\
\widehat{T}^{\,\sf cl}(z) & \sim &   \hskip .5cm \frac{\Delta_n}{z^2}
 \hskip 1.5cm  {\rm for}\;\; z \to \infty,
\end{eqnarray}
and consequently
\begin{equation}
\label{T:classical}
\widehat{T}^{\,\sf cl}(z)  =
Q^2 \sum\limits_{j=1}^{n-1}
\left[\frac{\delta_j}{(z-z_j)^2} + \frac{c_j}{z-z_j}\right].
\end{equation}
Combining  now (\ref{asymptotic:X}), (\ref{classical:withT})
and (\ref{T:classical})
with the conformal Ward identity \cite{Belavin:1984vu}
\begin{equation}
\label{Ward}
\left\langle \widehat{T}(z)\,C_{0,n}\right\rangle =
\sum\limits_{j=1}^{n-1}
\left[\frac{\Delta_j}{(z-z_j)^2} + \frac{1}{z-z_j}\frac{\partial}{\partial z_j}\right]
\left\langle \,C_{0,n}\, \right\rangle,
\end{equation}
we get the relation (\ref{PC}).
Amazingly, this relation obtained by general heuristic
path integral arguments turns out to provide an exact solution of
the  problem of the accessory parameters.
Indeed, it can be rigorously proved\footnote{For parabolic singularities
formula (\ref{PC}) has been proved by Takhtajan and Zograf. The details can be found
in \cite{TZ2}. In ref. \cite{ZoTa2} the extension of \cite{TZ2} to
compact Riemann surfaces has been presented. For general elliptic singularities eq. (\ref{PC})
has been proved in \cite{TZ} and non rigorously derived in \cite{Cantini:2001wr}.
It is also possible to construct the Liouville action satisfying
(\ref{PC}) for the so-called hyperbolic
singularities on the Riemann sphere, see \cite{Hadasz:2003kp}.}
that the formula
(\ref{PC}) yields the accessory parameters $c_j$
for which
the Fuchsian equation
\begin{equation}
\partial^{2}_{z}\psi(z) + \frac{1}{Q^2}\widehat{T}^{\,\sf cl}(z)\psi(z) = 0,
\end{equation}
admits a fundamental system of solutions
with  ${\sf PSU(1,1)}$ monodromies around all singularities. Note that
if $\{\chi_1(z), \chi_2(z)\}$ is such a system,
then the function $\phi(z,\bar z)$ determined by the relation
\begin{equation}
\label{phi}
\textrm{e}^{\phi(z,\bar z)}  =
{4\,|w'|^2 \over (1 - |w|^2 )^2},
\hskip 5mm
w(z) = {\chi_1(z)\over\chi_2(z)},
\end{equation}
satisfies (\ref{Liouville}) and (\ref{asymptot:elliptic}) (or
(\ref{asymptot:parabolic})).
The ${\sf PSU(1,1)}$ monodromy condition is then equivalent to the
existence of the
well defined hyperbolic metric on $C_{0,n}$.

\subsection{Fuchs equation with four elliptic/parabolic singularities}
Let us consider the case $n=4$ in which the four
elliptic/parabolic\footnote{In that case $\delta_1 = \ldots = \delta_4
  = \frac{1}{4}$.}
singularities are at the standard locations
$z_4 = \infty, z_3 = 1, z_2 = q, z_1 = 0$.
Accordingly, the expression of the classical energy-momentum tensor is
given by:
$$
T^{\,\sf cl}(z) = \frac{\delta_1}{z^2} + \frac{\delta_2}{(z-q)^2} + \frac{\delta_3}{(z-1)^2}
+ \frac{c_{1}(q)}{z} + \frac{c_{2}(q)}{z-q}+\frac{c_{3}(q)}{z-1}\;.
$$
and the first two relations of eq.~(\ref{dodatki2}) can be written as follows
$$
c_{1}(q)=\delta_1 + \delta_2 + \delta_3 - \delta_4 + (q-1) c_{2}(q),
\;\;\;\;\;\;\;\;\;\;\;
c_{3}(q)=\delta_4 - \delta_1 - \delta_2 - \delta_3 - q c_{2}(q).
$$
The Fuchsian differential equation (\ref{Fuchs}) has the following form:
\begin{equation}
\label{Fuchs2}
\partial^{2}_{z}\psi(z)+\left[\frac{\delta_1}{z^2}
+\frac{\delta_2}{(z-q)^2} +\frac{\delta_3}{(1-z)^2}
+\frac{\delta_1+\delta_2+\delta_3-\delta_4}{z(1-z)}
+\frac{q(1-q)c_{2}(q)}{z(z-q)(1-z)}\right]\psi(z)=0.
\end{equation}
There is only one undetermined accessory parameter, namely $c_{2}(q)$. This parameter
can be computed using the Polyakov conjecture once the classical four-point Liouville action
is known. Before considering that problem it is important to stress that equation (\ref{Fuchs2})
appears also in the context of the classical limit of DOZZ Liouville theory.

The partition function (\ref{partition}) corresponds in the
operator formulation to the correlation function of the primary
fields $V_{\alpha_j}(z_j,\bar z_j),$
\begin{equation}
\label{relation} \left\langle X \right\rangle  =
\Big\langle
V_{\alpha_n}(\infty,\infty)\ldots V_{\alpha_1}(z_1,\bar z_1)
\Big\rangle,
\end{equation}
with conformal weights
\[
\Delta_j = \alpha_j(Q-\alpha_j)
\]
where
\[
\alpha_j = \frac{Q}{2}\left(1 + \xi_j\right),
\;\;\;\;\;
Q = b + \frac{1}{b}.
\]
The DOZZ four-point correlation function for the standard locations
$z_4 = \infty, z_3 = 1, z_2 = q, z_1 = 0$
is expressed as an integral over the continuous spectrum
\begin{eqnarray}
\label{four:point:}
&&
\hspace*{-2.5cm}
\Big\langle
V_{\alpha_4}(\infty,\infty)V_{\alpha_3}(1,1)V_{\alpha_2}(q,\bar q)V_{\alpha_1}(0, 0)
\Big\rangle =
\\
\nonumber
&& \int\limits_{\frac{Q}{2} + i{\mathbb R}^{+}}\!\!\!\!\!\!\!d\alpha\;
C(\alpha_4,\alpha_3,\alpha)C(Q-\alpha,\alpha_2,\alpha_1) \left|
{\cal F}_{1+6Q^2,\Delta}\left[^{\Delta_3\ \Delta_2}_{\Delta_4\
\Delta_1}\right](q) \right|^2.
\end{eqnarray}
Let
\[
\mbox{\bf 1}_{\Delta,\Delta} = \sum_{I}
(|\,\xi_{\Delta,I}\,\rangle\otimes |\xi_{\Delta,I}\rangle)
(\langle\,\xi_{\Delta,I}\,|\otimes\langle\,\xi_{ \Delta,I}\,|)
\]
be an operator that projects onto the space spanned by the states
belonging to the conformal family with the highest weight $\Delta$. The
correlation function with the $\mbox{\bf 1}_{\Delta,\Delta} $
insertion factorizes into the product of the holomorphic and
anti-holomorphic factors,
\begin{eqnarray}
\label{c4}
&&
\hspace*{-1cm}
 \Big\langle
V_4(\infty,\infty)V_3(1,1)\mbox{\bf 1}_{\Delta,\Delta}
V_2(q,\bar q)V_1(0,0)
\Big\rangle  =  \\
 \nonumber
&& C(\alpha_4,\alpha_3,\alpha)\,C(Q-\alpha,\alpha_2,\alpha_1)\,
{\cal F}_{\!1+6Q^2,\Delta}
\!\left[_{\Delta_{4}\;\Delta_{1}}^{\Delta_{3}\;\Delta_{2}}\right]
\!(q)\,
{\cal F}_{\!1+6Q^2,\Delta}
\!\left[_{\Delta_{4}\;\Delta_{1}}^{\Delta_{3}\;\Delta_{2}}\right]
\!(\bar q).
\end{eqnarray}
Assuming a path integral representation of the left hand side, one
should expect
in the limit $b \to 0$, with all the weights being heavy,
i.~e. $\Delta,\Delta_i \sim \frac{1}{b^2}$,
the  following asymptotic behavior
\begin{equation}
\label{a4}
\Big\langle
V_4(\infty,\infty)V_3(1,1)\mbox{\bf 1}_{\Delta,\Delta}
V_2(q,\bar q)V_1(0,0)
 \Big\rangle \sim
{\rm e}^{-\frac{1}{b^2}
S^{\sf cl}_{\sf L}(\delta_i,q;\delta)
}.
\end{equation}
On the other hand, the $b\to 0$ limit of the DOZZ
coupling constants
\cite{Zamolodchikov:1995aa,Hadasz:2003he} gives as a result\footnote{
The quantity $S^{\sf cl}_{\sf L}(\delta_3,\delta_2,\delta_1)$ is
the classical three-point Liouville action whose form is known
for various types of singularities, see:
\cite{Zamolodchikov:1995aa,Hadasz:2003he}.
}
\begin{equation}
\label{asymptotC}
C(\alpha_4,\alpha_3,\alpha)C(Q-\alpha,\alpha_2,\alpha_1)
 \sim
{\rm e}^{-\frac{1}{b^2}\left(
S^{\sf cl}_{\sf L}(\delta_4,\delta_3,\delta)
+ S^{\sf cl}_{\sf L}(\delta,\delta_2,\delta_1)
\right)}.
\end{equation}
It follows that the conformal block should have the following asymptotic
behavior when $b\to 0$
\begin{equation}
\label{defccb}
{\cal F}_{\!1+6Q^2,\Delta}
\!\left[_{\Delta_{4}\;\Delta_{1}}^{\Delta_{3}\;\Delta_{2}}\right]
\!(q)
\; \sim \;
\exp \left\lbrace
\frac{1}{b^2}\,f_{\delta}
\!\left[_{\delta_{4}\;\delta_{1}}^{\delta_{3}\;\delta_{2}}\right]
\!(q)
\right\rbrace,
\end{equation}
so that
\begin{equation}
\label{deltaaction}
S^{\sf cl}_{\sf L}(\delta_i,q;\delta)=S^{\sf cl}_{\sf L}(\delta_4,\delta_3,\delta)
+ S^{\sf cl}_{\sf L}(\delta,\delta_2,\delta_1)-
f_{\delta}
\!\left[_{\delta_{4}\;\delta_{1}}^{\delta_{3}\;\delta_{2}}\right](q)
-\bar f_{\delta}
\!\left[_{\delta_{4}\;\delta_{1}}^{\delta_{3}\;\delta_{2}}\right](\bar q).
\end{equation}
It should be stressed that the asymptotic behavior (\ref{defccb}) is a nontrivial
statement concerning the quantum conformal block. Although there
is no proof of this property, it seems to be well confirmed together with its consequences by
sample numerical calculations \cite{HJP,P}.
One of such consequences is a relation between the classical four-point block
$f_{\delta}\!\left[_{\delta_{4}\;\delta_{1}}^{\delta_{3}\;\delta_{2}}\right]\!(q)$
and the accessory parameter $c_{2}(q)$ in eq. (\ref{Fuchs2}) as it will be
shown below.

Consider the null field corresponding to the null vector on the second level
of the Verma module. This is given by
\begin{equation}
\label{null} \chi_{-\frac{b}{2}}(z)=
\left[L_{-2}(z)-\frac{3}{2(2\Delta_{-\frac{b}{2}}+1)}\,L_{-1}^{2}(z)\right] V_{-\frac{b}{2}}(z,\bar
z),
\end{equation}
where $V_{\alpha=-\frac{b}{2}}$ is the degenerate primary
field with the degenerate weight
$\Delta_{-\frac{b}{2}}=-\frac{3}{4}\,b^2-\frac{1}{2}$.
It turns out that the
projected five-point correlation function on a sphere  with the null
field (\ref{null}) must vanish:
\begin{equation}
\label{chi} \left\langle\chi_{-\frac{b}{2}}(z) X\right\rangle_\Delta \; \equiv
\; \Big\langle V_4(\infty,\infty)V_3(1,1)\mbox{\bf
1}_{\Delta,\Delta} \chi_{-\frac{b}{2}}(z) V_2(q,\bar
q)V_1(0,0) \Big\rangle \;=\; 0.
\end{equation}
The above condition and the conformal Ward identities on the sphere
\cite{Belavin:1984vu} imply that the five-point function with the
degenerate operator $V_{-\frac{b}{2}}(z)$ satisfies the
equation:
\begin{eqnarray}
\label{Fuchs1} && \left[ \frac{\partial^2}{\partial z^2}
-b^2\left(\frac{1}{z} - \frac{1}{1-z}\right)\frac{\partial}{\partial
z} \right] \left\langle V_{-\frac{b}{2}}(z) X\right\rangle_\Delta =
 \\
\nonumber && -b^2 \left[\frac{\Delta_1}{z^2} +
\frac{\Delta_2}{(z-q)^2} +  \frac{\Delta_3}{(1-z)^2} +
\frac{\Lambda}{z(1-z)} + \frac{q(1-q)}{z(z-q)(1-z)}
\frac{\partial}{\partial q}\right]
\left\langle V_{-\frac{b}{2}}(z) X\right\rangle_\Delta,
\end{eqnarray}
where $ \Lambda\equiv\Delta_1 + \Delta_2 +\Delta_3
+\Delta_{-\frac{b}{2}}-\Delta_4.$
Let us assume that all the weights $\Delta$, $\Delta_i$, $i=1,\ldots,4$
in (\ref{chi}) and then in (\ref{Fuchs1}) are heavy, i.e.
$\Delta=\frac{1}{b^2}\,\delta$, $\Delta_i=\frac{1}{b^2}\,\delta_i$,
$\delta,\delta_i = {\cal O}(1)$.
In the limit $b \to 0$ only the operator with
weight $\Delta_{-\frac{b}{2}}$ remains light
($\Delta_{-\frac{b}{2}} = {\cal O}(1)$) and its presence in the
correlation function has no influence on the classical dynamics.
Then, for $b\to 0$
\begin{equation}
\label{a5}
\left\langle V_{-\frac{b}{2}}(z) X \right\rangle_\Delta \sim
\psi(z)\,
{\rm e}^{-{1\over b^2}\left(
S^{\sf cl}_{\sf L}(\delta_4,\delta_3,\delta) +
S^{\sf cl}_{\sf L}(\delta,\delta_2,\delta_1)
-f_{\delta}\!\left[_{\delta_{4}\;\delta_{1}}^{\delta_{3}\;\delta_{2}}\right](q)
-\bar f_{\delta}\!\left[_{\delta_{4}\;\delta_{1}}^{\delta_{3}\;\delta_{2}}\right](\bar q)\right)}
\end{equation}
and from (\ref{Fuchs1}) and (\ref{a5}) we get eq. (\ref{Fuchs2}) where the unknown accessory
parameter is given by
\begin{equation}
\label{accessory2}
c_2(q) =  {\partial\over \partial q}\,
f_{\delta}\!\left[_{\delta_{4}\;\delta_{1}}^{\delta_{3}\;\delta_{2}}\right]\!(q).
\end{equation}
The relation (\ref{accessory2}) is nothing but the Polyakov conjecture in the case
under consideration.

Indeed, in the semiclassical limit $b\to 0$ the left hand side
of formula (\ref{four:point:}) takes the form ${\rm
e}^{-\frac{1}{b^2} S_{\sf L}^{\,\sf
cl}(\delta_4,\delta_3,\delta_2,\delta_1;q)}$, where
we have used the shorthand notation
$$
S_{\sf L}^{\,\sf cl}(\delta_4,\delta_3,\delta_2,\delta_1;q) \equiv
S_{\sf L}^{\,\sf
cl}(\delta_4,\delta_3,\delta_2,\delta_1;\infty,1,q,0) .
$$
The right hand side of (\ref{four:point:}) is in this limit determined by the
saddle point approximation
$$
 {\rm e}^{-\frac{1}{b^2} S_{\sf L}^{\,\sf cl}(\delta_i,q)}
\; \approx \;
 \int\limits_0^\infty\!dp\; {\rm e}^{-\frac{1}{b^2} S_{\sf L}^{\,\sf cl}(\delta_i,q;\delta)}
$$
where $ \delta\equiv\delta_{s}(q)={\textstyle {1\over 4}} +p_{s}(q)^2
$ and the s-channel
saddle point Liouville momentum $p_s(q)$, is determined by the condition
\begin{equation}
\label{saddle} {\partial \over \partial p} S_{\sf L}^{\,\sf cl}
(\delta_i,q;{\textstyle {1\over 4}} +p^2)_{|p=p_s}=0\ .
\end{equation}
One gets thus the factorization
\begin{eqnarray}
\label{clasfact} S_{\sf L}^{\,\sf
cl}(\delta_4,\delta_3,\delta_2,\delta_1;q) &=& S_{\sf L}^{\,\sf
cl}(\delta_4,\delta_3,\delta_s(q)) +
S_{\sf L}^{\,\sf cl}(\delta_s(q),\delta_2,\delta_1)\nonumber\\[10pt]
&-&\,f_{\delta_s(q)}
\!\left[_{\delta_{4}\;\delta_{1}}^{\delta_{3}\;\delta_{2}}\right](q)
-\bar f_{\delta_s(q)}
\!\left[_{\delta_{4}\;\delta_{1}}^{\delta_{3}\;\delta_{2}}\right](\bar
q)
\end{eqnarray}
first obtained in \cite{Zamolodchikov:1995aa}. Having the classical four-point
Liouville action (\ref{deltaaction}), one can apply the Polyakov
conjecture and calculate the accessory
parameter:
\begin{eqnarray}\label{c2}
c_{2}(q)&=&-\frac{\partial}{\partial q}S_{\sf L}^{\,\sf cl}(\delta_i,q)\nonumber
\\[5pt]
&=&
-\frac{\partial}{\partial p}
S_{\sf L}^{\,\sf cl}(\delta_i,q,{\textstyle\frac{1}{4}}+p^2)\Big|_{p=p_{s}(q)}
\;\frac{\partial p_{s}(q)}{\partial q}-\frac{\partial}{\partial q}
S_{\sf L}^{\,\sf cl}(\delta_i,q,{\textstyle\frac{1}{4}}+p^2)\Big|_{p=p_{s}(q)}\nonumber
\\[5pt]
&=&
-\frac{\partial}{\partial q}
S_{\sf L}^{\,\sf cl}(\delta_i,q,{\textstyle\frac{1}{4}}+p^2)\Big|_{p=p_{s}(q)}
=
\frac{\partial}{\partial q}\,f_{\frac{1}{4}+p^2}
\!\left[_{\delta_{4}\;\delta_{1}}^{\delta_{3}\;\delta_{2}}\right]\!(q)\Big|_{p=p_{s}(q)}.
\end{eqnarray}
Here the saddle point equation (\ref{saddle}) and the factorization (\ref{clasfact})
have been used.

Hence, the problem of computing the accessory parameter
$c_{2}(q)$ is equivalent to the problem of calculating the classical four-point
block. The function $f_{\delta}\!\left[_{\delta_{4}\;\delta_{1}}^{\delta_{3}\;\delta_{2}}\right](q)$
is known in general only as a formal power series with coefficients
calculated exploiting the asymptotic
behavior (\ref{defccb}) and the expansion of the quantum conformal
block (see the appendix).
However, one can sum up the series defining the classical four-point block by
applying the 'chiral' part of the AGT correspondence. More concretely,
one should apply its 'classical version', which relates the classical
limit of conformal blocks
to the Nekrasov--Shatashvili limit of the Nekrasov instanton partition
functions. The derivation of the analytic expression for the four-point classical block
will be one of our main tasks in the next section.

\section{Accessory parameters from gauge theory}
\subsection{Nekrasov--Shatashvili limit}
Consider the instanton part of the Nekrasov partition function of the
${\cal N}=2$
supersymmetric ${\sf U(2)}$ gauge theory with four hypermultiplets in the fundamental
representation \cite{N}:
\begin{eqnarray}\label{inst}
{\cal Z}_{\sf inst}^{{\sf U(2)}, N_f = 4} &=&
1+\sum\limits_{k=1}^{\infty}\frac{q^k}{k!}
\left(\frac{\epsilon_1+\epsilon_2}{\epsilon_1 \epsilon_2}\right)^k\,{\cal Z}_k
\nonumber
\\
&=&
1+\sum\limits_{k=1}^{\infty}\frac{q^k}{k!}
\left(\frac{\epsilon_1+\epsilon_2}{\epsilon_1 \epsilon_2}\right)^k\,
\oint\frac{d\phi_1}{2\pi i}\ldots\oint\frac{d\phi_k}{2\pi i}\;\Omega_k,
\end{eqnarray}
where
\begin{eqnarray*}
\Omega_k &=& \prod\limits_{I=1}^{k}
\frac{\prod\limits_{\alpha=1}^{4}\left(\phi_I + m_{\alpha}\right)}
{\prod\limits_{u=1}^{2}\left(\phi_I - a_u - i0 \right)\left(\phi_I - a_u + \epsilon_1 + \epsilon_2 + i0\right)}
\\
&\times &
\prod_{\begin{array}{c}\scriptstyle I,J=1\\[-7pt]\scriptstyle I \neq J \end{array}}^{k}
\frac{\left(\phi_I - \phi_J\right)\left(\phi_I - \phi_J +\epsilon_1 +\epsilon_2\right)}
{\left(\phi_I - \phi_J + \epsilon_1 + i0\right)\left(\phi_I - \phi_J + \epsilon_2 + i0\right)}.
\end{eqnarray*}
We will assume that $a_u,\epsilon_1,\epsilon_2\in \mathbb{R}$.
The contours in (\ref{inst}) go over the real axis and close in the
upper half-plane.
Recall, that the poles which contribute to (\ref{inst}) are
in correspondence with  pairs
of Young diagrams  $Y=\lbrace Y_{1}, Y_{2} \rbrace$:
\begin{equation}
\label{poles}
Y \To \phi_I = \phi_{u,r,s} = a_u + (r-1)\epsilon_1 + (s-1)\epsilon_2,
\;\;\;\;\;\;\;\;\;\;
u=1, 2.
\end{equation}
The index $r$ labels the columns while $s$
runs over the rows of the diagram $Y_u$. The parameters $\epsilon_1$ and
$\epsilon_2$ describe the size
of the box $(r,s)\in Y_u$ in the horizontal and vertical direction
respectively.
The total number of boxes $|Y| = |Y_1|+|Y_2|$ is equal to the instanton
number $k$.
The instanton sum
 over $k$ in (\ref{inst}) can be rewritten as a sum over a pairs
of Young diagrams as follows:
$$
{\cal Z}_k = \sum_{Y\atop |Y|=k} {\cal Z}_Y.
$$
The contributions ${\cal Z}_Y$ to the instanton sum correspond to
those obtained
by performing (in some specific order) the contour integrals in (\ref{inst}).

Now we want to calculate the limit $\epsilon_2 \to 0$ of the instanton
partition function (\ref{inst}).
Based on the arguments developed in \cite{NekraOkun}, it is reasonable
to expect that
for vanishingly small values of $\epsilon_2$ the dominant contribution
to the instanton partition function
(\ref{inst}) will occur when
$k\sim \frac{1}{\epsilon_2}$.
\footnote{This statement becomes evident in the trivial case in which
${\cal Z}_k = 1$ $\forall k=1,2,\ldots$. For $\epsilon_2 \to 0$
and $x=q/\epsilon_2 \in \mathbb{R}_{>0}$ we have then from
eq.~(\ref{inst})
\begin{eqnarray*}
{\cal Z}_{\sf inst} &=& \sum\limits_{k=0}^{\infty}\frac{1}{k!}
\left(\frac{q}{\epsilon_2}\right)^k
=\sum\limits_{k=0}^{\infty}\frac{x^k}{k!}\,
= \textrm{e}^x = \frac{\textrm{e}^{x\log x}}{\textrm{e}^{x\log x - x}}
\sim\frac{\textrm{e}^{x\log x}}{\textrm{e}^{\log x!}}=\frac{x^{x}}{x!}.
\end{eqnarray*}
This means that the whole sum is dominated by a single term with
$k\sim x\to \infty$.
Unfortunately, we have found no proof of that mechanism in the general case.}
For future purposes it will be necessary to compute
the leading behavior of $\log \left|q^k \Omega_k
\right|$ for large $k$
(i.~e. small values of $\epsilon_2$ and finite $\epsilon_1$).
After simple calculations \footnote{The computations which lead
to (\ref{leading}) are elementary
and rely on the Taylor expansion of $\log(x\pm\epsilon_2)$ for small
$\epsilon_2$, i.e.:
$\log(x\pm\epsilon_2)=\log(x)\pm\frac{\epsilon_2}{x}+O(\epsilon_{2}^{2})$.}
we find:
\begin{eqnarray}\label{leading}
\log\left| q^k \Omega_k \right| &\sim &
\frac{1}{\epsilon_2}\Big[
\epsilon_2 k\log |\,q|
+\epsilon_2 \sum\limits_{I=1}^{k} \left[
\sum\limits_{\alpha=1}^{4}\log\left|\phi_I + m_{\alpha}\right|
-\sum_{u=1}^{2}\log\left(\left|\phi_I - a_u \right|
\left|\phi_I - a_u + \epsilon_1\right|\right)\right]\nonumber
\\
&+&
\epsilon_{2}^{2}\!\!\!
\sum\limits_{\begin{array}{c}\scriptstyle I,J=1\\[-7pt]\scriptstyle I \neq J \end{array}}^{k}
\left[\frac{1}{\phi_I - \phi_J + \epsilon_1} - \frac{1}{\phi_I -
    \phi_J}\right] \Big].
\end{eqnarray}
In eq.~(\ref{leading}) it is implicitly understood that the poles
$\phi_I$ are obtained from eq.~(\ref{poles})  in the
limit $\epsilon_2\to 0$.
It turns out that the right hand side of eq.~(\ref{leading})
is equal up to the factor
$\frac{1}{\epsilon_2}$
to the instantonic free energy
$
{\cal H}^{{\sf U(2)},N_f = 4}_{{\sf inst}}$.
Note that in the limit $\epsilon_2\to 0$ the poles
form a continuous distribution:
\begin{equation}\label{bieguny}
\phi_I = \phi_{u,r}\in\left[x_{u,r}^{0}, x_{u,r}\right]
\end{equation}
where
\begin{eqnarray*}
x_{u,r}^{0} &=& a_u + (r-1)\epsilon_1,
\;\;\;\;\;\;\;\;\;\;
u=1,2,
\;\;\;\;\;\;\;\;\;\;
r=1,\ldots,\infty,
\\
x_{u,r} &=& a_u + (r-1)\epsilon_1 +\omega_{u,r}.
\end{eqnarray*}
In the language of Young diagrams the two formulas above can be
explained as follows.
When $\epsilon_2$ is very small, the number of boxes $k_{u,r}$ in the vertical
direction (the number of rows) is very large, while the quantity
$\omega_{u,r}=\epsilon_2 k_{u,r}$ is
expected to be finite. In other words, we obtain a continuous distribution of
rows in the limit under consideration. As a consequence, in order to
evaluate the instanton free energy,
the sums 'over the instantons' in (\ref{leading}) may be replaced by
continuous integrals in the row index, with the range of
integration specified by eq.~(\ref{bieguny}). It is thus possible to write:
\begin{equation}\label{sumI}
\epsilon_2\sum\limits_{I}\To\sum\limits_{u,r}
\int\limits_{x_{u,r}^{0}}^{x_{u,r}}d\phi_{u,r}.
\end{equation}
The integration limits $x_{u,r}^{0}$ and $x_{u,r}$ represent
the bottom and the top ends of the $r$-th column in $Y_u$ respectively.
Applying eq.~(\ref{sumI}) to eq.~(\ref{leading}) one gets
\begin{eqnarray}\label{H}
{\cal H}^{{\sf U(2)},N_f = 4}_{{\sf inst}}\left(x_{u,r}\right) &=&
\sum\limits_{u,v=1}^{2} \sum\limits_{r,l=1}^{\infty}
\Big[-F\left(x_{u,r}-x_{v,l}+\epsilon_1\right) + F\left(x_{u,r}-x_{v,l}^{0}+\epsilon_1\right)
\\
&&\hspace{-70pt}+\;
F\left(x_{u,r}^{0}-x_{v,l}+\epsilon_1\right) - F\left(x_{u,r}^{0}-x_{v,l}^{0}+\epsilon_1\right)
+ F\left(x_{u,r}-x_{v,l}\right)
\nonumber\\
&&\hspace{-70pt}-\;
F\left(x_{u,r}-x_{v,l}^{0}\right) - F\left(x_{u,r}^{0}-x_{v,l}\right)
+ F\left(x_{u,r}^{0}-x_{v,l}^{0}\right)
\Big]
\nonumber\\
&&\hspace{-110pt}+ \sum\limits_{u,v=1}^{2} \sum\limits_{r=1}^{\infty}
\Big[ -F\left(x_{u,r}-a_v\right) +F\left(x_{u,r}^{0}-a_v\right)
-
F\left(x_{u,r}-a_v +\epsilon_1\right)+F\left(x_{u,r}^{0}-a_v +\epsilon_1\right)\Big]
\nonumber\\
&&\hspace{-110pt}+ \sum\limits_{u=1}^{2}\sum_{r=1}^{\infty}\sum_{\alpha=1}^{4}
\Big[ F\left(x_{u,r}+m_\alpha\right) - F\left(x_{u,r}^{0}+m_\alpha\right)\Big]
+ \sum\limits_{u=1}^{2}\sum\limits_{r=1}^{\infty}
\left(x_{u,r}-(r-1)\epsilon_1 - a_u\right)\log |\,q|\;,\nonumber
\end{eqnarray}
where $F(x)=x(\log |\,x| - 1)$.

Let us turn to the main problem of our interest.
According to the ideology of \cite{NekraOkun} the Nekrasov instanton
partition function in the limit $\epsilon_2\to 0$ can be represented as
follows:
\begin{equation}\label{path}
{\cal Z}_{\sf inst}^{{\sf U(2)},N_f=4} \sim\int \left[\prod\limits_{u,r}dx_{u,r}\right]
\,\exp\left\lbrace{\frac{1}{\epsilon_2}\,
{\cal H}_{\sf inst}^{{\sf U(2)},N_f=4}(x_{u,r})}\right\rbrace,
\end{equation}
where the integral is over the infinite set of variables
$\left\lbrace x_{u,r} : u=1,2; r=1,\ldots,\infty\right\rbrace$.
As a consequence, the Nekrasov-Shatashvili limit
of ${\cal Z}_{\sf inst}^{{\sf U(2)},N_f=4}$ is nothing
but the critical value of ${\cal H}_{\sf inst}^{{\sf U(2)},N_f=4}$:
\begin{equation}\label{critvalue}
{\cal W}_{\sf inst}^{{\sf U(2)},N_f=4}\;\equiv \;
\lim\limits_{\epsilon_2 \to 0}\epsilon_2 \log {\cal Z}_{\sf inst}^{{\sf U(2)},N_f=4}
\;=\;{\cal H}_{\sf inst}^{{\sf U(2)},N_f=4}(x_{u,r}^{*}),
\end{equation}
where $x_{u,r}^{*}$ denotes the 'critical configuration' extremizing
the 'free energy' (\ref{H}).


\subsection{Saddle point equation}
The extremality condition for the 'action' ${\cal H}_{\sf inst}^{{\sf U(2)},N_f=4}$
given by (\ref{H}) reads as follows:
$$
\left|\,q\left(\prod_{v=1}^{2}\prod_{l=1}^{\infty}
\frac{(x_{u,r}-x_{v,l}-\epsilon_1)(x_{u,r}-x_{v,l}^{0}+\epsilon_1)}
{(x_{u,r}-x_{v,l}+\epsilon_1)(x_{u,r}-x_{v,l}^{0}-\epsilon_1)} \right)
\left(\frac{\prod\limits_{\alpha=1}^{4}(x_{u,r}+m_{\alpha})}{\prod\limits_{v=1}^{2}
(x_{u,r}-a_{v})(x_{u,r}-a_{v}+\epsilon_1)}
\right)\right|=1.
$$
This implies that either the following identity:
\begin{equation}
\label{saddle2}
-q\left(\prod_{v=1}^{2}\prod_{l=1}^{\infty}
\frac{(x_{u,r}-x_{v,l}-\epsilon_1)(x_{u,r}-x_{v,l}^{0}+\epsilon_1)}
{(x_{u,r}-x_{v,l}+\epsilon_1)(x_{u,r}-x_{v,l}^{0}-\epsilon_1)} \right)
\left(\frac{\prod\limits_{\alpha=1}^{4}(x_{u,r}+m_{\alpha})}{\prod\limits_{v=1}^{2}
(x_{u,r}-a_{v})(x_{u,r}-a_{v}+\epsilon_1)}
\right)=1
\end{equation}
or its analog in which $-q$ is replaced by $+q$ are holding.
To find the solution of eq.~(\ref{saddle2}) will be the main task of
this subsection.
Eq. (\ref{saddle2}) can be regularized assuming that there is an integer
$L$ such that the length of the column $\omega_{u,r}$ is equal to
zero for $r>L$.
Analyzing eq.~(\ref{saddle2}) in such a case, i.e. when $l=1,\ldots, L$,
one can observe that the column lengths extremizing the 'free energy'
are of the order
$\omega_{u,r}\sim {\cal O}(q^r)$. For example, at order $q^L$ one can write
\begin{equation}\label{solution}
x_{u,r}^{*}\equiv x_{u,r} = a_u + (r-1)\epsilon_1 +\omega_{u,r}(\,q)
=a_u + (r-1)\epsilon_1 +\sum\limits_{n=r}^{L}\omega_{u,r,n}\;q^n.
\end{equation}
Here the symbols $\omega_{u,r,n}$ denote the contributions to the
coefficients $\omega_{u,r}$ at the $n-$th order in $q$.
Now it is possible to
 solve equation (\ref{saddle}) starting from $L=1$ and deriving
 recursively
the  $\omega_{u,r,n}$'s step by step up to desired order.
The calculation of the first few coefficients  $\omega_{u,r,n}$
is presented below.
\begin{enumerate}
\item
For $L=1$ the equation (\ref{saddle2}) becomes
\begin{eqnarray*}
&&\hspace{-45pt} -q \prod_{v=1}^{2}\left(
\frac{(x_{u,r}-x_{v,1}-\epsilon_1)(x_{u,r}-x_{v,1}^{0}+\epsilon_1)}
{(x_{u,r}-x_{v,1}^{0}-\epsilon_1)
(x_{u,r}-x_{v,1}+\epsilon_1)
(x_{u,r}-a_v)(x_{u,r}-a_v +\epsilon_1)}\right)
\prod\limits_{\alpha=1}^{4}(x_{u,r}+m_\alpha) = 1
\end{eqnarray*}
or equivalently
\begin{eqnarray}\label{saddle4}
&& q (x_{u,r}-x_{1,1}-\epsilon_1)(x_{u,r}-x_{2,1}-\epsilon_1)
(x_{u,r}+m_1)(x_{u,r}+m_2)
(x_{u,r}+m_3)(x_{u,r}+m_4)\nonumber\\
&+&
(x_{u,r} - a_1 - \epsilon_1)(x_{u,r} - a_2 - \epsilon_1)
(x_{u,r}- x_{1,1}+\epsilon_1)
(x_{u,r}- x_{2,1}+\epsilon_1)\nonumber
\\
&\times &
(x_{u,r}- a_1)(x_{u,r}- a_2) \;=\; 0.
\end{eqnarray}
Hereafter we fix the freedom in the choice of the parameters $a_1$ and
$a_2$ by setting $(a_1, a_2)=(a, -a)$. Thus,
expanding (\ref{solution}) up to the first order in
$q$ and substituting the result into (\ref{saddle4}) one finds that
\begin{eqnarray}\label{omegaU2}
{\omega}_{1,1,1} = -\frac{\prod\limits_{\alpha=1}^{4}(a + m_\alpha)}{\epsilon_1 2a (2a + \epsilon_1)},
\;\;\;\;\;\;\;\;
{\omega}_{2,1,1} = -\frac{\prod\limits_{\alpha=1}^{4}(a - m_\alpha)}{\epsilon_1 2a (2a - \epsilon_1)}.
\end{eqnarray}
\item
For $L=2$ the system of linear equations obtained from (\ref{saddle2})
yields the second order corrections to the length of the first column.
The coefficients are
\begin{eqnarray*}
\label{omega112}
{\omega}_{1,1,2} &=&
\Big\lbrace
(a+m_1)(a+m_2)(a+m_3)(a+m_4)
\left(a-\epsilon_1\right)\nonumber
\\
&\times & \Big( 8a^5 (a+m_1)(a+m_2)(a+m_3)(a+m_4)
- a^2\epsilon_{1}^{6}\left( 2a+{\sf m}\right) - a^2\epsilon_{1}^{7}\nonumber
\\
&+& \epsilon_{1}^{5} \left[13a^4 + 4a^3 {\sf m} + 2a^2 \mu + a {\hat \mu}-\mathfrak{m}\right]\nonumber
\\
&+& a \epsilon_{1}^{4} \left[5a^4 + 6a^3 {\sf m} -7a^2 \mu + 2a {\hat \mu} - 3\mathfrak{m}\right]\nonumber
\\
&-& a^2\epsilon_{1}^{3} \left[51a^4 +5a^3 {\sf m} + 19a^2 \mu - 11a {\hat \mu} + 3\mathfrak{m}\right]\nonumber
\\
&+& 2a^3\epsilon_{1}^{2} \left[ -11a^4 + 3a^3 {\sf m} -3a^2 \mu + 3a {\hat \mu} - 11\mathfrak{m}\right]\nonumber
\\
&+& 4a^4\epsilon_1 \left[7a^4 + 5a^3 {\sf m} + 3a^2 \mu + a {\hat \mu} - \mathfrak{m}\right]
\Big)\Big\rbrace
\\
&\times &
\Big\lbrace 8a^3\epsilon_{1}^{3}\left(2a+\epsilon_1 \right)^2 \left(2a^2 + a\epsilon_1 - \epsilon_{1}^{2}\right)
\left(4a^3 - 4a^2\epsilon_1 - a\epsilon_{1}^{2}+\epsilon_{1}^{3}\right)\Big\rbrace^{-1},
\end{eqnarray*}
and
\begin{eqnarray*}
{\omega}_{2,1,2} &=&
\Big\lbrace
(a - m_1)(a - m_2)(a - m_3)(a - m_4)
\left(a  + \epsilon_1\right)
\\
&\times & \Big( 8a^5 (a-m_1)(a-m_2)(a-m_3)(a-m_4)
+ a^2\epsilon_{1}^{6}\left( -2a+{\sf m}\right) + a^2\epsilon_{1}^{7}
\\
&+& \epsilon_{1}^{5} \left[-13a^4 + 4a^3 {\sf m} - 2a^2 \mu + a {\hat \mu} + \mathfrak{m}\right]
\\
&+& a \epsilon_{1}^{4} \left[5a^4 - 6a^3 {\sf m} -7a^2 \mu + 2a {\hat \mu} - 3\mathfrak{m}\right]
\\
&+& a^2\epsilon_{1}^{3} \left[51a^4 - 5a^3 {\sf m} + 19a^2 \mu + 11a {\hat \mu} + 3\mathfrak{m}\right]
\\
&-& 2a^3\epsilon_{1}^{2} \left[11a^4 + 3a^3 {\sf m} + 3a^2 \mu + 3a {\hat \mu} + 11\mathfrak{m}\right]
\\
&-& 4a^4\epsilon_1 \left[7a^4 - 5a^3 {\sf m} + 3a^2 \mu - a {\hat \mu} - \mathfrak{m}\right]
\Big)\Big\rbrace
\\
&\times &
\Big\lbrace 8a^3\epsilon_{1}^{3}\left(\epsilon_1 -2a\right)^2 \left(-2a^2 + a\epsilon_1 + \epsilon_{1}^{2}\right)
\left(-4a^3 - 4a^2\epsilon_1 + a\epsilon_{1}^{2}+\epsilon_{1}^{3}\right)\Big\rbrace^{-1},
\end{eqnarray*}
where
$$
{\sf m}\equiv\sum\limits_{i=1}^{4}m_i,
\;\;\;\;\;\;\;\;
\mathfrak{m}\equiv\prod\limits_{i=1}^{4}m_i,
\;\;\;\;\;\;\;\;
\mu\equiv\sum\limits_{1\leq i<j\leq 4} m_{i}m_{j},
\;\;\;\;\;\;\;\;
{\hat \mu}\equiv\sum\limits_{1\leq i<j<k\leq 4} m_{i}m_{j}m_{k}.
$$
Moreover, from eq.~(\ref{saddle2}) with $L=2$ it is also possible to
determine the length of the second column at the leading
order in $q^2$. Indeed, one can derive the following coefficients:
\begin{eqnarray*}
{\omega}_{1,2,2} &=&
- \frac{\prod\limits_{\alpha=1}^{4}(a + m_\alpha)(a + \epsilon_1 + m_\alpha)}
{8a\epsilon_{1}^{3}(a+\epsilon_1)(2a + \epsilon_{1})^2},
\\
{\omega}_{2,2,2} &=&
- \frac{\prod\limits_{\alpha=1}^{4}(a - m_\alpha)(a - \epsilon_1 - m_\alpha)}
{8a\epsilon_{1}^{3}(a-\epsilon_1)(\epsilon_{1}-2a)^2}.
\end{eqnarray*}
\end{enumerate}


\subsection{Twisted superpotential, classical block and accessory parameter}
Knowing the extremal lengths of the columns one can calculate the critical value
of the 'free
energy' (\ref{critvalue}), i.e. the so-called twisted superpotential.
In order to compute this critical value it is convenient first to
calculate the derivative of
${\cal W}_{\sf inst}^{{\sf U(2)},N_f=4}(\,q, a, m_i; \epsilon_1)$ with
respect to $q$:
\begin{eqnarray}\label{derW}
\frac{\partial}{\partial q} {\cal W}_{\sf inst}^{{\sf U(2)},N_f=4}(\,q, a, m_i; \epsilon_1)
&=&
\frac{\partial {\cal H}_{\sf inst}}{\partial x_{u,r}}\frac{\partial x_{u,r}}{\partial q}
+
\frac{\partial {\cal H}_{\sf inst}}{\partial q}
=
\frac{1}{q}\sum\limits_{u,r}\omega_{u,r}.
\end{eqnarray}
In the above calculation we have used the fact that $\partial {\cal
  H}_{\sf inst}/\partial x_{u,r} =0$.
It is easy to realize that the last term
 in (\ref{derW}) coincides with the sum over the column lengths of the
 'critical' Young diagram.
Performing this sum one  obtains the correct expansion
of the twisted superpotential. Indeed, using (\ref{derW}) one gets:
\begin{eqnarray}
q\frac{d}{dq}\,{\cal W}_{\sf inst}^{{\sf U(2)},N_f=4}
&=&
\sum\limits_{r}\left(\omega_{1,r}(\,q)+\omega_{2,r}(\, q)\right)
=
\sum\limits_{r}\left[\sum\limits_{n=r}\left(\omega_{1,r,n} + \omega_{2,r,n}\right)q^n\right]\nonumber
\\[8pt]
&=&
\left[\left(\omega_{1,1,1} + \omega_{2,1,1}\right) q
+\left(\omega_{1,1,2} + \omega_{2,1,2}\right) q^2 +\ldots \right]\nonumber
\\[8pt]
&+&
\left[\left(\omega_{1,2,2} + \omega_{2,2,2}\right) q^2
+\left(\omega_{1,2,3} + \omega_{2,2,3}\right) q^3 +\ldots\right]
+
\ldots\;.
\end{eqnarray}
Then,
\begin{eqnarray}
\label{Wexp}
{\cal W}_{\sf inst}^{{\sf U(2)},N_f=4}
&=&
\left({\omega}_{1,1,1}+{\omega}_{2,1,1}\right)q +
\left({\omega}_{1,1,2}+{\omega}_{2,1,2}+{\omega}_{1,2,2}+{\omega}_{2,2,2}\right)\frac{q^2}{2}
+\ldots\;\nonumber
\\[8pt]
&=& {\cal W}_{1}^{{\sf U(2)},N_f=4}\,q + {\cal W}_{2}^{{\sf U(2)},N_f=4}\,q^2 +\ldots\;.
\end{eqnarray}
The expansion (\ref{Wexp}) with the coefficients calculated from the
saddle point equation exactly agrees with
that obtained directly form the expansion of the instanton partition function.
Moreover, assuming the following relations between parameters:
\begin{eqnarray}
\label{etas}
m_1 = \epsilon_1 \left(\eta_1 + \eta_2 - \frac{1}{2}\right),
&&
m_2 = \epsilon_1 \left(\eta_2 - \eta_1 + \frac{1}{2}\right),\nonumber
\\
m_3 = \epsilon_1 \left(\eta_3 + \eta_4 - \frac{1}{2}\right),
&&
m_4 = \epsilon_1 \left(\eta_3 - \eta_4 + \frac{1}{2}\right),
\;\;\;\;\;\;\;\;
a = \epsilon_1 \left(\eta -\frac{1}{2}\right)
\end{eqnarray}
and using the expression of the coefficients
$\omega_{u,r,n}$'s calculated in the previous paragraph, one can check that
\begin{eqnarray}
\frac{1}{\epsilon_1}\,{\cal W}_{1}^{{\sf U(2)},N_f=4} &=&
\frac{1}{\epsilon_1}\,\left({\omega}_{1,1,1}+{\omega}_{2,1,1}\right)=
\frac{(\delta + \delta_2 - \delta_1)(\delta + \delta_3 - \delta_4) - 4\delta \eta_2 \eta_3}{2\delta}\nonumber
\\
&=&
\textsf{f}^{\,1}_{\delta}\!\left[_{\delta_{4}\;\delta_{1}}^{\delta_{3}\;\delta_{2}}\right] - 2\eta_2 \eta_3
= \textsf{f}^{\,1}_{\delta}\!\left[_{\delta_{4}\;\delta_{1}}^{\delta_{3}\;\delta_{2}}\right] -
\frac{(m_1 + m_2)(m_3 + m_4)}{2\epsilon_{1}^{2}},\label{wuone}
\end{eqnarray}
and
\begin{eqnarray}
\frac{1}{\epsilon_1}\,{\cal W}_{2}^{{\sf U(2)},N_f=4} &=&
\frac{1}{\epsilon_1}\,\frac{1}{2}
\left({\omega}_{1,1,2}+{\omega}_{2,1,2}+{\omega}_{1,2,2}+{\omega}_{2,2,2}\right)
\nonumber\\
&=&
\textsf{f}^{\,2}_{\delta}\!\left[_{\delta_{4}\;\delta_{1}}^{\delta_{3}\;\delta_{2}}\right] - \eta_2 \eta_3
=\textsf{f}^{\,2}_{\delta}\!\left[_{\delta_{4}\;\delta_{1}}^{\delta_{3}\;\delta_{2}}\right] -
\frac{1}{2}\left(2\eta_2 \eta_3\right),\label{wutwo}
\end{eqnarray}
where
\begin{eqnarray}
\label{classW}
\delta = \eta\left(1-\eta\right),
\;\;\;\;\;\;\;\;\;\;\;\;
\delta_i = \eta_{i}\left(1-\eta_{i}\right),
\;\;\;\;\;\;\;\;\;\;\;\;
i=1,\ldots,4.
\end{eqnarray}
In eqs.~(\ref{wuone}) and (\ref{wutwo}) the symbols
$\textsf{f}^{\,n}_{\delta}\!\left[_{\delta_{4}\;\delta_{1}}^{\delta_{3}\;\delta_{2}}\right]$'s
for $n=1,2$ are the first two coefficients of the classical four-point
block introduced in (\ref{defccb}) (see appendix):
\begin{eqnarray}
\label{classblock}
f_{\delta}\!\left[_{\delta_{4}\;\delta_{1}}^{\delta_{3}\;\delta_{2}}\right]\!(\,q)
&=&
(\delta-\delta_1-\delta_2) \log q +
{\sf f}_{\delta}\!\left[_{\delta_{4}\;\delta_{1}}^{\delta_{3}\;\delta_{2}}\right]\!(\,q)
\nonumber\\
&=& (\delta-\delta_1-\delta_2) \log q +  \sum_{n=1}^\infty
{\sf f}^{\,n}_{\delta}\!\left[_{\delta_{4}\;\delta_{1}}^{\delta_{3}\;\delta_{2}}\right]\,q^n.
\end{eqnarray}
It is thus reasonable to expect that
\begin{eqnarray}\label{Wvsf1}
\frac{1}{\epsilon_1}\,\left({\cal W}_{1}^{{\sf U(2)},N_f=4}\,q
+ {\cal W}_{2}^{{\sf U(2)},N_f=4}\,q^2 +\ldots\;\right)
&=&
\left(\textsf{f}^{\,1}_{\delta}\!\left[_{\delta_{4}\;\delta_{1}}^{\delta_{3}\;\delta_{2}}\right]
- 2\eta_2 \eta_3\right)\,q\nonumber
\\
&&\hspace{-100pt}+
\left(\textsf{f}^{\,2}_{\delta}\!\left[_{\delta_{4}\;\delta_{1}}^{\delta_{3}\;\delta_{2}}\right]
- \frac{1}{2}\left(2\eta_2 \eta_3\right)\right)\,q^2 + \ldots\;.
\end{eqnarray}
The conjectured identity (\ref{Wvsf1}) is nothing but the expansion of
both sides of the relation:
\begin{eqnarray}
\label{ClassAGTsphere}
\frac{1}{\epsilon_1}\;
{\cal W}_{{\sf inst}}^{{\sf U(2)},N_f=4}(q,a,m_i;\epsilon_1)
&=&
{\sf f}_{\delta}
\!\left[_{\delta_{4}\;\delta_{1}}^{\delta_{3}\;\delta_{2}}\right](q)
+\frac{(m_1+m_2)(m_3+m_4)}{2\epsilon_{1}^{2}}
\;\log(1-q).
\end{eqnarray}
The identities (\ref{Wvsf1}) or (\ref{ClassAGTsphere}) are justified not only by the above calculations.
Note that the relation (\ref{ClassAGTsphere}) is nothing else but the
classical/Nekrasov--Shatashvili limit of the
AGT relation\footnote{For a proof of the AGT relation on $C_{0,n}$ see
  \cite{AlbaFatLitTarnp}.}:
\begin{eqnarray*}\label{AGTsphere}
q^{\Delta_1 + \Delta_2 -\Delta}\,
{\cal F}_{c,\Delta}\!\left[_{\Delta_{4}\;\Delta_{1}}^{\Delta_{3}\;\Delta_{2}}\right]\!(\,q)
&=& (1-q)^{-\frac{(m_1+m_2)(m_3+m_4)}{2\epsilon_1\epsilon_2}}\,
\mathcal{Z}^{{\sf U(2)}, N_f = 4}_{{\sf inst}}(q,a,m_i;\epsilon_1, \epsilon_2),
\end{eqnarray*}
where
\begin{eqnarray}
\label{external}
c=1+6\frac{(\epsilon_1+\epsilon_2)^2}{\epsilon_1\epsilon_2}
\equiv 1+6Q^2,&\;\;\;\;\;&
\Delta=\frac{(\epsilon_1+\epsilon_2)^2-4a^2}{4\epsilon_1\epsilon_2},\nonumber\\
\Delta_1=
\frac{\frac{1}{4}(\epsilon_1\!+\!\epsilon_2)^2-\frac{1}{4}(m_1\! - \!m_2)^2}{\epsilon_1\epsilon_2},
&\;\;\;\;\;&
\Delta_2=
\frac{\frac{1}{2}(m_1\!
+\! m_2)(\epsilon_1\!+\!\epsilon_2
-\frac{1}{2}(m_1\! +\! m_2))}{\epsilon_1\epsilon_2},\nonumber
\\
\Delta_3=
\frac{\frac{1}{2}(m_3\! +\! m_4)
(\epsilon_1\!+\!\epsilon_2\!-\!\frac{1}{2}(m_3 \!+\! m_4))}{\epsilon_1\epsilon_2}
,
&\;\;\;\;\;&
\Delta_4=
\frac{\frac{1}{4}(\epsilon_1\!+\!\epsilon_2)^2\!-\!\frac{1}{4}(m_3\! -\! m_4)^2}{\epsilon_1\epsilon_2}\nonumber
\end{eqnarray}
and
$$
Q = b+\frac{1}{b} \equiv
{\sqrt{\frac{\epsilon_2}{\epsilon_1}}}
+\sqrt{\frac{\epsilon_1}{\epsilon_2}}
\;\;\;\;\;\;\;\;\;\;
\Longleftrightarrow
\;\;\;\;\;\;\;\;\;\;
b = {\sqrt{\frac{\epsilon_2}{\epsilon_1}}}.
$$

As a final conclusion of this subsection let us write down the two
main results of the present work. The first one is
a novel representation for the four-point classical block
with four elliptic/parabolic external classical weights
and a hyperbolic intermediate classical weight.
Indeed, from eqs.~(\ref{critvalue}), (\ref{classblock}) and (\ref{ClassAGTsphere})
we have
\begin{eqnarray}
\label{classblockNEW}
f_{\delta}\!\left[_{\delta_{4}\;\delta_{1}}^{\delta_{3}\;\delta_{2}}\right]\!(\,q)
&=& (\delta-\delta_1-\delta_2) \log q -
\frac{(m_1+m_2)(m_3+m_4)}{2\epsilon_{1}^{2}}
\;\log(1-q) \nonumber
\\
&+& \frac{1}{\epsilon_1}\,{\cal H}_{\sf inst}^{{\sf
    U(2)},N_f=4}(x_{u,r}^{*}(\,q))\nonumber
\\
&=& (\delta-\delta_1-\delta_2) \log q -
\frac{(m_1+m_2)(m_3+m_4)}{2\epsilon_{1}^{2}}
\;\log(1-q) \nonumber
\\
&+& \frac{1}{\epsilon_1}\,{\cal W}_{\sf inst}^{{\sf U(2)},N_f=4}(\,q, a, m_i; \epsilon_1),\label{fourpointcb}
\end{eqnarray}
where the classical conformal weights are parameterized as in (\ref{classW})
with $\eta$'s given by (\ref{etas}).
Knowing the classical four-point block from (\ref{classblockNEW}) and
applying eqs.~(\ref{c2}) and
(\ref{derW}), one arrives at the following expression of the accessory
parameter $c_2(q)$:
\begin{eqnarray}
c_{2}(q) \;=\; \frac{\delta-\delta_1-\delta_2 + \frac{1}{\epsilon_1}
\sum_{u,r}\omega_{u,r}(q, a, m_i;\epsilon_1)}{q}+
\frac{2\eta_2 \eta_3}{1-q},\label{accpar}
\end{eqnarray}
The four masses $m_i$ appearing in eq.~(\ref{accpar})
are given by eq.~(\ref{etas}) and the
vacuum expectation value
$
a=-i\epsilon_1 p_{s}(q)
$
is proportional to the s-channel momentum $p_{s}(q)$.
Hence, we have found that the accessory parameter
$c_{2}(q)$ is related to the sum of column lengths of the 'critical'
Young diagram.
The latter can be rewritten using the contour integral representation.
Following \cite{Pogho1,Fucito} let us define the  functions:
\begin{eqnarray*}
Y(z) &=& \prod\limits_{u=1}^{2}\exp\left\lbrace\frac{z}{\epsilon_1}\,
\psi\left(\frac{a_u}{\epsilon_1}\right)\right\rbrace\prod_{r=1}^{\infty}
\left( 1-\frac{z}{x_{u,r}} \right)\exp\left\lbrace \frac{z}{x_{u,r}^{0}} \right\rbrace,
\\
Y_{0}(z) &=& \prod\limits_{u=1}^{2}\exp\left\lbrace\frac{z}{\epsilon_1}\,
\psi\left(\frac{a_u}{\epsilon_1}\right)\right\rbrace\prod_{r=1}^{\infty}
\left( 1-\frac{z}{x_{u,r}^{0}} \right)\exp\left\lbrace \frac{z}{x_{u,r}^{0}} \right\rbrace,
\end{eqnarray*}
where $\psi(z)=\partial_{z}\log\Gamma(z)$. The functions $Y(z)$, $Y_{0}(z)$ are holomorphic
with zeros located at $x_{u,r}$ and $x_{u,r}^{0}$ respectively. Then,
\begin{eqnarray*}
\sum_{u,r}\limits\omega_{u,r} &=&  \left[\sum_{u,r}\left(x_{u,r}-x_{u,r}^{0}\right)\right]\Big|_{x_{u,r}=x_{u,r}^{*}}
=
\left(\oint\limits_{\gamma}\frac{dz}{2\pi i}\,z\partial_{z}\log\frac{Y(z)}{Y_{0}(z)}\right)\Big|_{x_{u,r}=x_{u,r}^{*}},
\end{eqnarray*}
where $\gamma$ encloses all the points $x_{u,r}$, $x_{u,r}^{0}$, $u=1,2$, $r=1,\ldots,\infty$.

\section{Conclusions}
The original results of the present paper are:
\begin{itemize}
\item
The derivation of
the generic classical four-point block
provided in eq.~(\ref{fourpointcb}), where the classical four-point
block has been written in terms of the
critical value of the
instanton 'free energy' ${\cal H}_{\sf inst}^{{\sf U(2)},N_f=4}$.
\item
The
derivation in closed form of the accessory parameter $c_2(q)$
appearing in the Fuchs equation with four
parabolic/elliptic singularities.
So far the expression of this accessory parameter was unknown.
From eq.~(\ref{accpar})   $c_2(q)$  can be interpreted
as the sum of all column
lengths of the 'critical'
Young diagram which extramizes ${\cal H}_{\sf inst}^{{\sf U(2)},N_f=4}$.
\end{itemize}

The above results possess interesting further applications.
According to the formula (\ref{pullback}) the solution of the accessory
parameters problem for the Fuchs equation
on $C_{0,4}$ offers the possibility of constructing the solution of the
Liouville equation on $C_{0,4}$. In the case of a sphere with
four parabolic singularities the above results pave the way
for the construction of the uniformization map and the computation of
the so-called geodesic length functions \cite{HJP}.

As a next point let us discuss the possible extensions of this work.
One of its aims was to find an analytic expression of
the classical four-point block exploiting the 'chiral' sector of the
AGT correspondence on $C_{0,4}$. It has been found that the
classical block can be expressed in terms of the ${\sf U(2)}$, $N_f = 4$
instanton twisted superpotential. If the classical four-point block
and the three-point classical Liouville action are available
one can construct the four-point classical action (cf. (\ref{deltaaction})).
The three-point Liouville action can be recovered from the DOZZ
structure constant
in the classical limit. As a consequence, due to the AGT duality the
four-point classical
action should correspond to the full effective twisted superpotential
${\cal W}^{{\sf U(2)},N_f=4} =
{\cal W}^{{\sf U(2)},N_f=4}_{\sf pert} +
{\cal W}^{{\sf U(2)},N_f=4}_{\sf inst}$.
Based on this example it is reasonable to expect that:
\begin{enumerate}
\item
the ${\sf U(2)}$ Nekrasov partition function
with $N_f = n$ flavors encodes an information on
the $n$-point classical Liouville action $S_{\sf L}^{{\sf cl}, (n)}$
on the sphere;
\item
the classical Liouville action $S_{\sf L}^{{\sf cl},(n)}$ can be recovered from the
Nekrasov partition function in the Nekrasov--Shatashvili limit;
\item
the classical action $S_{\sf L}^{{\sf cl},(n)}$ can be expressed as the critical
value of the gauge theory 'free energy'.
\end{enumerate}
If these conjectures are true they provide a direct way to calculate
$S_{\sf L}^{{\sf cl},(n)}$.

In this work we have studied a version of the problem of accessory parameters which is related
to the classical Liouville theory on the sphere. Let us stress that there exists also an
analogous problem  in the case of the torus topology
\cite{Menotti1,Menotti2}.
It would be interesting to investigate whether the AGT correspondence
can be applied
to solve the problem of accessory parameters also in that case.

Finally, it seems to be an interesting
task to study possible overlaps of our results and those in papers
\cite{Teschner,NekraRosSha}.

\appendix
\section{Quantum and classical four-point conformal blocks}
Let
\begin{eqnarray*}
{\cal V}_{\Delta_j} &=& \bigoplus_{n=0}^{\infty}{\cal V}_{\Delta_j}^{n},
\\
{\cal V}_{\Delta_j}^{n} &=& \textrm{Span}\Big\lbrace
\nu^n_{\Delta_j, I}=L_{-I}\nu_{\Delta_j} = L_{-i_{k}}\ldots
L_{-i_{2}}L_{-i_{1}} \nu_{\Delta_j}
\\
&&\hspace{60pt}:
I=( i_{k}\geq \ldots\geq i_{1}\geq 1)
\;\textrm{an}\; \textrm{ordered} \;
\textrm{set}\; \textrm{of }\;\textrm{positive} \;\textrm{integers} \;
\\
&&\hspace{180pt}
\textrm{of}\;
\textrm{the}\;\textrm{length}\;|I|\equiv i_{1}+\ldots+i_{k}=n
\Big\rbrace
\end{eqnarray*}
be the Verma module with the highest weight state $\nu_{\Delta_j}$.
The chiral vertex operator is the linear map
$$
V{^{\Delta_3}_\infty}{^{\Delta_2}_{\:z}}{^{\Delta_1}_{\;0}} :
{\cal V}_{\Delta_2} \otimes {\cal V}_{\Delta_1}
\To
{\cal V}_{\Delta_3}
$$
such that for all $\xi_2 \in {\cal V}_{\Delta_2}$ the operator
$$
V(\xi_2 | z) \equiv
V{^{\Delta_3}_\infty}{^{\Delta_2}_{\:z}}{^{\Delta_1}_{\;0}}
(\xi_2\otimes\,\cdot\,):{\cal V}_{\Delta_1}
\To
{\cal V}_{\Delta_3}
$$
satisfies the following conditions
\begin{eqnarray}
\label{CVO}
\left[L_n , V\!\left(\nu_{2}|z\right)\right] &=& z^{n}\left(z
\frac{d}{dz} + (n+1)\Delta_2
\right) V\!\!\left(\nu_{2}|z\right)\,,\;\;\;\;\;\;\;\;n\in\mathbb{Z}
\\
V\!\!\left(L_{-1}\xi_{2}|z\right) &=& \frac{d}{dz}V\!\!\left(\xi_{2}|z\right),\nonumber
\\
V\!\!\left(L_{n}\xi_{2}|z\right) &=& \sum\limits_{k=0}^{n+1}
\left(\,_{\;\;k}^{n+1}\right)
(-z)^{k}\left[L_{n-k}, V\!\!\left(\xi_{2}|z\right)\right]\,,
\;\;\;\;\;\;\;\;n>-1,\nonumber
\\
V\!\!\left(L_{-n}\xi_{2}|z\right) &=&\sum\limits_{k=0}^{\infty}
\left(\,_{\;\;n-2}^{n-2+k}\right)
z^k\,L_{-n-k}\,V\!\!\left(\xi_{2}|z\right)\nonumber
\\
&+& (-1)^{n}\sum\limits_{k=0}^{\infty}
\left(\,_{\;\;n-2}^{n-2+k}\right)
z^{-n+1-k}\,\,V\!\!\left(\xi_{2}|z\right)\,L_{k-1},
\;\;\;\;\;n>1\nonumber
\end{eqnarray}
and
$$
\left\langle\nu_{\Delta_3},V\!\!\left(\nu_{2}|z\right)\nu_{\Delta_1}\right\rangle
\;=\;z^{\Delta_3 -\Delta_{2}-\Delta_1}.
$$
Let $q$ be the moduli of the 4-punctured sphere. The quantum four-point conformal block
is defined as the formal power series:
\begin{eqnarray}
\label{block} {\cal
F}_{c,\Delta}\!\left[_{\Delta_{4}\;\Delta_{1}}^{\Delta_{3}\;\Delta_{2}}\right]\!(\,q)
&=& q^{\Delta-\Delta_{2}-\Delta_{1}}\left( 1 + \sum_{n=1}^\infty
{\cal
F}^{\,n}_{c,\Delta}\!\left[_{\Delta_{4}\;\Delta_{1}}^{\Delta_{3}\;\Delta_{2}}\right]
q^{\,n} \right)
\end{eqnarray}
with coefficients given by
\begin{eqnarray}\label{coef}
{\cal
F}^{\,n}_{c,\Delta}\!\left[_{\Delta_{4}\;\Delta_{1}}^{\Delta_{3}\;\Delta_{2}}\right]
&=& \label{blockCeef} \sum\limits_{n=|I|=|J|}
\left\langle\nu_{\Delta_4},V(\nu_3|1)\nu_{\Delta,I}\right\rangle
\;
\Big[G_{c,\Delta}\Big]^{IJ}
\;
\left\langle\nu_{\Delta,J},V(\nu_2|1)\nu_{\Delta_1}\right\rangle.
\end{eqnarray}
Above $\Big[ G_{c,\Delta}\Big]^{IJ}$ is the
inverse of the Gram matrix
$ \Big[G_{c,\Delta}\Big]_{IJ}=\langle \nu_{\Delta,I},\nu_{\Delta,J} \rangle$
of the standard symmetric bilinear form in the Verma module.
Taking into account the covariance properties (\ref{CVO}) of the primary chiral
vertex operator with respect to the Virasoro algebra one can calculate
the matrix elements in (\ref{coef}). Hence, for lower orders of the expansion the
coefficients (\ref{coef}) can be easily computed directly from definition.
For instance,
\begin{eqnarray*}
{\cal F}^{\,1}_{c,\Delta}\!\left[_{\Delta_{4}\;\Delta_{1}}^{\Delta_{3}\;\Delta_{2}}\right]
&=&\frac{(\Delta+\Delta_3-\Delta_4)(\Delta+\Delta_2-\Delta_1)}{2\Delta},
\\
{\cal F}^{\,2}_{c,\Delta}\!\left[_{\Delta_{4}\;\Delta_{1}}^{\Delta_{3}\;\Delta_{2}}\right]
&=&
\Big[
\left(4\Delta(1+2\Delta)\right)^{-1}
(\Delta-\Delta_1 +\Delta_2)
(1+\Delta-\Delta_1+\Delta_2)
\\
&\times &
(\Delta+\Delta_3 -\Delta_4)
(1+\Delta+\Delta_3-\Delta_4)
\\
&+&
\left(
\Delta-\Delta^{2}-\Delta_1-\Delta_2
+3(\Delta_2 -\Delta_1)^2 - 2\Delta (\Delta_1 +\Delta_2)
\right)
\\
&\times &
\left(
\Delta-\Delta^{2}-\Delta_3 -\Delta_4
+3(\Delta_3 -\Delta_4)^2 - 2\Delta (\Delta_3 +\Delta_4)
\right)
\Big]
\\
&\times &
\left[
2(1+2\Delta)^2 \left( c-\frac{4\Delta(5-8\Delta)}{2+4\Delta}\right)
\right]^{-1}.
\end{eqnarray*}
As the dimension of ${\cal V}^n_\Delta$ grows rapidly with $n$,
the calculations of conformal block coefficients by inverting
the Gram matrices become very laborious for higher orders.
A more efficient method based on recurrence relations for the coefficients
can be used \cite{Zam,Zamolodchikov:ie,ZZ}.

Let us assume that all the conformal weights in the conformal block
are heavy. Then, the asymptotic behavior (\ref{defccb}) implies
the following expansion of the 4-point classical block:
\begin{eqnarray}
\label{classblock2}
f_{\delta}\!\left[_{\delta_{4}\;\delta_{1}}^{\delta_{3}\;\delta_{2}}\right]\!(\,q)
&=& (\delta-\delta_1-\delta_2) \log q +  \sum_{n=1}^\infty
q^{n}\, {\sf f}^{\,n}_{\delta}\!\left[_{\delta_{4}\;\delta_{1}}^{\delta_{3}\;\delta_{2}}\right]
\\
&=& (\delta-\delta_1-\delta_2) \log q + \lim\limits_{b \to 0} {b^2} \log\left(1 +
\sum_{n=1}^\infty {\cal
F}^{\,n}_{\!c,\Delta}\!\left[_{\Delta_{4}\;\Delta_{1}}^{\Delta_{3}\;\Delta_{2}}\right]
q^{\;n} \right)\nonumber.
\end{eqnarray}
The coefficients
${\sf f}^{\,n}_{\delta}\!\left[_{\delta_{4}\;\delta_{1}}^{\delta_{3}\;\delta_{2}}\right]$
in (\ref{classblock2})
are calculated directly from the limit (\ref{defccb}) and the power expansion
of the quantum block. For example, expanding the logarithm into power series
and then taking the limit of each term separately for $n=1,2$ one finds
\begin{eqnarray*}
{\sf f}^{\,1}_{\delta}\!\left[_{\delta_{4}\;\delta_{1}}^{\delta_{3}\;\delta_{2}}\right]
&=&
\frac{(\delta + \delta_3 -\delta_4)(\delta + \delta_2 - \delta_1)}{2\delta},
\\
{\sf f}^{\,2}_{\delta}\!\left[_{\delta_{4}\;\delta_{1}}^{\delta_{3}\;\delta_{2}}\right]
&=&
\Big[16\delta^3 (4\delta +3)\Big]^{-1}
\Big[
13\delta^5 +\delta^4 \left(18\delta_2 -14\delta_1 +18\delta_3 -14\delta_4 + 9\right)
\\
&+&
\delta^3 \Big(
\delta_{1}^{2}+\delta_{2}^{2}-2\delta_1 (\delta_2 +6\delta_3 -10\delta_4 +6)
\\
&+&
4\delta_2 (5\delta_3 -3\delta_4 +3) + (\delta_3 -\delta_4)(\delta_3 -\delta_4 +12)
\Big)
\\
&-&
3\delta^2
\Big(\delta_{1}^{2}(2\delta_3 +2\delta_4 -1)
+ 2\delta_1 (\delta_{3}^{2}+\delta_{4}^{2} +2\delta_3 +\delta_2 -2\delta_2 \delta_3 - 2\delta_4 (\delta_2+\delta_3 +1))
\\
&+&
\delta_{2}^{2}(2\delta_3 + 2\delta_4 -1)
+2\delta_2 (\delta_3 -\delta_4 -2)(\delta_3 -\delta_4)-(\delta_3 -\delta_4)^2
\Big)
\\
&+&
5\delta (\delta_1 -\delta_2)^2 (\delta_3 -\delta_4)^2 - 3(\delta_1 -\delta_2)^2 (\delta_3 -\delta_4)^2
\Big].
\end{eqnarray*}

\acknowledgments
The authors are grateful to Marco Matone for stimulating questions and comments.
Special thanks go to Leon A. Takhtajan for remarks and very valuable advices
concerning the first version of the paper.

This research has been supported in part by the Polish National
Science Centre under Grant No. N202 326240.

\end{document}